\documentstyle[12pt]{article}
\begin{document}
\begin{titlepage}
\begin{center}
{\Large\bf Nucleon number dependence of longitudinal radii in ion-ion
collisions as a signature of the onset of collective expansion}
\end{center}
\vspace{1cm}
\begin{center}
{\large J\'an Pi\v{s}\'ut${}^{a)b)}$, Neva Pi\v{s}\'utov\'a${}^{b)}$
and Petr Z\'avada${}^{c)}$}
\end{center}
\vspace{1cm}
\begin{center}
  {\it a)Laboratoire de Physique Corpusculaire,Universit\'{e} Blaise Pascal,
  Clermont-Ferrand,
  F-63177 Aubi\`{e}re,Cedex,France \\
  b) Department of Physics,Comenius University,
  SK-842 15 Bratislava,Slovakia \\
  c)Institute of Physics,Czech Acad.Sci. Na Slovance 2, CZ-18040, Praha-8
    Czech Republic}
\end{center}
\vspace{1.5cm}

\abstract
  {In an attempt to disentangle the effects of nuclear geometry from 
   those of expansion we study the dependence of the longitudinal
   radius $(R_L)$ on nuclear numbers of colliding ions within a simple
   model of multiple nucleon- nucleon collisions.The model has built in 
   the nuclear geometry but there is no dynamical expansion.In principle,
   what is significantly beyond models of this type can be considered as
   caused by collective expansion.Taking into account the present errors
   of data, oversimplified model and the absence of accurate information
   on $R_L$ for pp interactions it is difficult to make strong
   statements about collisions of induced by light ions. 
   On the other hand
    Pb-Pb data give $R_L$ significantly higher than a geometrical
   picture could accomodate, presenting an evidence for the onset of
   collective expansion,or of onset of a new dynamical regime
    between O-Au and Pb-Pb collisions.The situation with
   S-Au and S-Pb collisions is less clear because of a rather large
   difference of data on $R_L$ in these two cases.}
\end{titlepage}
\section{Introduction}
\label{intro}

Heavy ion collisions provide the only way to laboratory studies of dense
hadronic matter and hopefully also of theoretically predicted
Quark- Gluon Plasma $(QGP)$. Hanbury- Brown and Twiss $(HBT)$ interferometry
[1,2]  brings information about dimensions of the homogeneity region
[3,4,5] of particle production.

Space-time evolution of proton-nucleus $(pA)$ and of ion-ion $(AB)$
collisions depends on details of dynamics and on values of important
parameters,like the formation time of secondary hadrons,which are both
only partially understood.It is therefore impossible to predict
accurately when the collective expansion of hadronic matter or of $QGP$
sets on.In an ideal situation dynamics of heavy-ion collisions would be 
known up to a few parameters which could be determined by HBT interferometry.
Unfortunately the dynamics is unknown and because of that we shall start
with simplest assumptions possible.

We suppose that in the CERN SPS energy range collisions of protons and of
light ions with nuclear targets consist of multiple nucleon-nucleon
collisions,with dynamics close to that described in Ref.[6] . For
earlier work in this direction, see Ref.[7] .In each
of collisions,except for the last one, nucleon loses some fraction of
its energy.In the last one it fragments roughly like in a $pp$ collision.
In those kinematical regions where the production of secondary particles
is dominated by last collisions of individual nucleons,the model of Ref.
[6] is close to predictions of the "wounded" nucleon [8] or
"participating" nucleon [9] models.The model of Ref.[6] admits
a parton model interpretation,according to which fast partons,including
valence quarks,do not participate in individual nucleon-nucleon collisions,
what makes the picture consistent [10] with the experimental
information on the A,B dependence of the Drell-Yan dilepton production.
This model is not exactly that of multiple nucleon-nucleon scattering in
the whole rapidity domain.For secondary hadrons in the central rapidity
region the picture might be close to that of independent nucleon- nucleon
collisions,since slow hadrons in the nucleon- nucleon cms might be formed
in between of successive collisions.

The purpose of the present note is to discuss a formula for the dependence
of $R_L$ on nucleon numbers $A$ and $B$ of colliding ions,based on the 
picture of multiple nucleon-nucleon collisions.The formula might roughly
correspond also to other models,provided that the collective expansion
is rather small.
 We shall then compare the 
formula with  available data [11-14] of NA-44, NA-35 and
WA-93 Collaborations obtained at the CERN-SPS and try to see whether the
data can be understood only by nuclear geometry and fragmentation of
nucleons or whether some dynamical expansion is required.We shall use two
approaches.In the former one we {\it assume} 
 that  expanding hadronic matter is {\it not} present in proton and
light ion collisions with nuclei in the CERN SPS energy region.The
parameters in the model are then determined from this condition.
Having
parameters fixed in this way we compare the formula
with data on $R_L$ obtained in heavy ion collisions in particular in
$Pb+Pb$ and $S+Pb$ ones.If values of $R_L$ in these collisions are
not larger than what follows from such an extrapolation of data from
lower A's and B's then also $S+Pb$ and $Pb+Pb$ collisions would be
most likely  dominated
just by nuclear geometry.If,on the other hand,longitudinal radii as
measured in $S+Pb$ and $Pb+Pb$ are larger than such extrapolations,then
a new dynamical mechanism,presumably caused by the collective expansion
of matter, must be responsible for the increase of $R_L$.

 In the latter approach we take the model literally and calculate the
parameters.Deviations from the model are supposed to give evidence for
the presence of expansion.

In both approaches we are trying to  find a threshold at which
 $R_L$ "decouples" from  simple
geometrical dependence.Any point of decoupling or a point when one type
of behaviour of $R_L$ changes to another one is most likely related to the
onset of a new dynamics of the collision.A threshold for "decoupling"
from a simple geometrical dependence signals most probably a transition to
a collective expansion of matter.

  Note that such thresholds should be correlated also with changes
of patterns of other signatures,like enhancement of dilepton 
and photon production,
increased $J/\psi$ suppression,appearance of azimutal asymmetries in
noncentral collisions,etc.

\section{Dependence of $R_L$ on A and B in a model of multiple nucleon- 
 nucleon collisions}
\label{geometry}
The correlation function for two identical pions with momenta 
$\vec k_1$ and $\vec k_2$ is written in the standard way
\begin{equation}
C(\vec k_1,\vec k_2)=1+{\lambda \left | \int e^{i(\vec k_1-\vec k_2).\vec r}
\rho (\vec r;\vec K)d^3\vec r \right |}^2
\label{eq1}
\end{equation}
where $\vec K = (\vec k_1 +\vec k_2)/2 $ and $\rho (\vec r;\vec K)$ is the
density of sources of the corresponding homogeneity region [3,4,5].
To simplify the discussion we have assumed that both pions have the same
energy.In that case $\rho (\vec r;\vec K)$ in Eq.(1) is  
integral over time
 of the space-time density distribution $\rho (\vec r,t;\vec K)$.
The density of sources $\rho (\vec r;\vec K)$ is asummed to be given, in
some approximation, as
\begin{equation}
\rho (\vec r;\vec K) = {\frac {1}{{R_T}^2{R_L}{(2\pi )^{3/2}}}}
\exp \left (-\frac {z^2}{2{R_L}^2} -\frac {x^2 +y^2}{2{R_T}^2} \right )
\label{eq2}
\end{equation}
The Fourier transform corresponding to $\vec k_1 - \vec k_2 \equiv \vec q$
parallel to the beam $(\equiv z)$ axis becomes
\begin{equation}
\tilde \rho (\vec q) = exp(-\frac {{R_L^2}{q^2}}{2})
\label{eq3}
\end{equation}
and the correlation function in Eq.(1) is
\begin{equation}
C(q)=1+\lambda exp(-{R_L^2}{q^2})
\label{eq4}
\end{equation}
The longitudinal radius $R_L$ is related to the mean squared value of 
$z$ by
\begin{equation}
{R_L}^2 =  \langle z^2 \rangle = \int {z^2}\rho(\vec r;\vec K){d^3}\vec r
\label{eq5}
\end{equation}
We shall further assume that in individual nucleon-nucleon collisions
pions with small $z$-component of momentum (in the nucleon-nucleon c.m.s.)
are produced in a way similar to the one in collisions of free nucleons.The
process goes pressumably via formation and decay of resonances in both 
cases.We thus assume that the density of sources for identical pions in a
given nucleon-nucleon collision which took place in the point with
coordinate $\bar z$ is given as
\begin{equation}
{\rho _{nn}}(z-\bar z;\vec K)={\frac {1}{R_L(pp){(2{\pi })^{1/2}}}}
\exp -\left (-\frac {(z-\bar z)^2}{2{R_L}^2{(pp)}} \right )
\label{eq6}
\end{equation}
Here $R_L{(pp)}$ is the longitudinal radius as obtained by HBT 
interferometry in $pp$ collisions.The experimental information obtained
by the AFS Collaboration at the CERN ISR [15] indicates that $R_L{(pp)}$
is a bit larger than $1fm$.This result has been corroborated by further
evidence obtained in less direct measurements [16] in studies of
directional dependence of HBT interferometry,based on the method proposed
in Ref.[17].
The density of nucleon-nucleon collisions is parametrized as
\begin{equation}
{\rho }_l{(\bar z)} = {\frac {1}{l\sqrt 2\pi }}\exp (-\frac {{\bar z}^2}
{2l^2} )
\label{eq7}
\end{equation}
Where $l$ is given by Eq.(6).
The total density of sources is then given as
\begin{equation}
\rho (z;\vec K) = \int {\rho _{nn}}(z-\bar z;\vec K){\rho _l(z-\bar z)}
d\bar z
\label{eq8}
\end{equation}
where we have suppressed variables $x,y$.The Fourier transform of a convolution
is the product of Fourier transforms and this leads via Eqs.(1,7,8,9) to
\begin{equation}
C(q;\vec K)=1+\lambda \exp \left ( -(l^2(A,B) +R_L^2(pp))q^2
 \right )
\label{eq9}
\end{equation}
 and finally to
\begin{equation}
R_L^2(A,B)=R_L^2(p,p)+l^2(A,B)
\label{eq10}
\end{equation}
Where $l^2(A,B)$ is given in Eq.(7) and can be calculated as
\begin{equation}
 l^2(A,B)=<{\bar z}^2>
\label{eq11}
\end{equation} 
The interpretation of Eq.(10) is simple.The experimentally observable
longitudinal radius in $AB$ collisions consists of two parts.The former,
$R_L(pp)$ is due to resonance decays or to equivalent dynamical
reasons (production of clusters,etc.) and the latter is due to the geometry
of the $AB$ collision.

\section{Two approaches to the comparison of the formula with data}
\label{comparison}
We shall now compare the formula in Eq.(10) with data within the two
approaches discussed in the Introduction.

The {\bf former approach} is more qualitative.We assume that
in $A+B$ collision individual nucleon-nucleon interactions occur
in a region with longitudinal dimension
\begin{equation}
l \approx c{r_0}{\frac {1}{\gamma }} \left ( {(A-1)}^{1/3}+{(B-1)}^{1/3}
\right )
\label{eq12}
\end{equation}
Here $r_0 \approx 1.2fm$,$\gamma $ is the Lorentz contraction factor,
$c$ is a constant of the order $1$ and replacements $A \rightarrow (A-1)$,
$B \rightarrow (B-1)$ are unimportant for the data we shall discuss,but
they facilitate the transition to $pp$ collisions. 
 The constant $c$ may
include some details of dynamics as well as the fact that the approximation
$R_A \approx 1.2{r_0}{A^{1/3}}$ underestimates [18] radii of light
ions.
Conventions used correspond to those of NA-44 and we have recalculated the
NA-35 data accordingly.

 We shall consider the interference of identical 
pions with $K_z \approx 0$ in the c.m.s. of nucleon-nucleon collisions.
In this frame incoming
nuclei appear contracted by the Lorentz factor $\gamma = {s^{1/2}}/{2M}$
where $s$ is the square of the nucleon-nucleon c.m.s. energy and $M$ is
the nucleon mass.For $E_L=200AGeV$,$\gamma \approx 10$ and for $E_L=
160AGeV$ we have $\gamma \approx 9$.

From Eqs.(10) and (12) we find

\begin{equation}
{R_L}^2(A,B)={R_L}^2(pp)+\left ( (c{r_0}/\gamma)({(A-1)}^{1/3}+
{(B-1)}^{1/3}) \right )^2
\label{eq13}
\end{equation}
For $A=B=1$ we obtain ${R_L}(1,1)={R_L}(p,p)$ as we should.

The comparison of data on $R_L(A,B)$ as obtained at the CERN SPS with
Eq.(13) is presented in Fig.1.The data corresponding to $AB$ collisions
were taken from the following sources: $S+C$ from Ref.[11], $S+S$ from
Ref.[11]; $p+Pb$ from Ref.[13]; $S+Cu$ from Ref.[11];
$S+Ag$ from Ref.[11]; $O+Au$ from Ref.[11]; $S+Au$ from Ref.[11]; $S+Pb$
from Ref.[13] and $Pb+Pb$ from Refs.[13,14].
 
The straight line in Fig.1 corresponds to the following values of 
parameters entering Eq.(11)
\begin{equation}
{R_L}(pp) \approx 1.92 fm;\qquad \frac {c{r_0}}{\gamma } \approx 0.36 fm
\label{eq14}
\end{equation}
Since we have read the data of Ref.[13] from graphs and since the
statistics is continuously increasing we do not give errors of the
fit of data by Eq.(11),pointing out only that the determination
of ${R_L}(pp)$ by the data is less accurate than that of $c{r_0}/\gamma $.

The qualitative, and in our opinion the most important point, is
easily visible
in Fig.1. The data from $S+C$ up to $O+Au$ are compatible
with the straight line whereas ${R_L}(PbPb)$ is certainly much larger
than what could be ascribed to the linear dependence obtained from
proton- and light ion collisions with nuclear targets.We interpret this
as an evidence for the presence of a threshold occuring between $O+Au$
and $Pb+Pb$ collisions at the CERN SPS energies.Fig.1 contains also an
indication of a possible discrepancy between data on $R_L$ on $S+Au$
as given in Ref.[11] and on $S+Pb$ presented in Ref.[13]. Difference
between nucleon numbers of $Au(197)$ and $Pb(207)$ is hardly responsible
for a large difference between the corresponding values of $R_L$.
If the true value of ${R_L}{(SPb)}$ is closer to Ref.[13] the threshold
is probably around $S+Pb$ collisions.

The value of ${R_L}(pp) \approx 1.92 fm$ seems to be  larger than
expected,but since this is not very accurately determined by the "fit"
we shall not try to analyze possible reasons for that.

Taking $\gamma \approx 10$ and ${r_0} \approx 1.2 fm$ we find $c \approx
3$ what is also somewhat large.
The value of $c$ might be
also influenced by the underestimate of radii of light ions by the
expression $R_A = {r_0}A^{1/3}$,see Ref.[18].

Note that in the region where data in Fig.1 are presented the
dependence as given by Eq.(11) is not much different from a linear
relationship between ${R_L}(AB)$ and $(A-1)^{1/3} +(B-1)^{1/3}$.

Arguments given above are admittedly rather crude.In spite of simplifications
used we shall make an attempt to interpret the results.The experimental
value [13,14] of the square of longitudinal radius for $Pb+Pb$ collisions
is about $40 fm^2$,whereas the extrapolation from proton- and light ion
nuclear collisions gives about $20 fm^2$.The expansion should be responsible
for the difference.Putting ${R^2}_{L,exper} \approx {R^2}_{L,geom} +
{R^2}_{expan}$ we get $R_{expan} \approx 6fm $.Since $\langle z^2 \rangle 
={{R_L}^2} $ we have ${{\langle z^2 \rangle }^{1/2}}_{expan} \approx
5.6fm$.Assuming that for pions with $K_z \approx 0$ the homogeneity region
corresponds roughly to rapidity region $ -1/2 < y < 1/2 $ than the time of
expansion is roughly $ t_{expan} \approx {{ \langle z^2 \rangle }^{1/2}_{expan}}
/v(y=1/2) $ where $v(y=1/2)=tanh(1/2)$ is the velocity corresponding to
$y=1/2$.In this way we get the estimate $t_{expan} \approx 11.2fm/c$.

The {\bf latter approach} is more rigid. Values of $l^2(A+B)$ entering
Eqs.(10) and (11) are obtained from  nuclear geometry without any
free constants.We have calculated values of $l^2(A+B)$ for different
colliding nuclei by using the model of Ref.[19].In this model distribution
of nucleons in nuclei is given by a standard Woods- Saxon density. The
resulting values of $l^2(A+B)$ for central collisions in the c.m.s of
nucleon- nucleon interactions and taking into account Lorentz contraction
are as follows (in $fm^2$)
$$ l^2(S+C)=0.014 \qquad l^2(S+S)=0.018 \qquad l^2(S+Cu)=0.022 $$
\begin{equation}
l^2(S+Ag)=0.029 \qquad l^2(O+Au)=0.037 \qquad l^2(S+Au)=0.041 
\label{eq15}
\end{equation}
$$l^2(p+Pb)=0.038 \qquad l^2(S+Pb)=0.042 \qquad l^2(Pb+Pb)=0.055$$
In these calculations we have not taken into account deceleration of
nucleons by collisions.

Very similar values of $l^2(A+B)$ are obtained in a simple model in
which nucleus with radius R is replaced by a cylinder of length 2L 
with $L={\sqrt {3/5}} R$.This relation is given by the requirement that
$<z^2>$ is the same in both cases.

Values in Eq.(15) are rather small with respect to $R^2_L(pp)$ and the
corresponding curve in Fig.1 is practically flat.To draw the curve we 
need the value of $R^2_L(pp)$.
 The interpretation
of data is thus
 made difficult by the absence of experimental information on
$R^2_L(pp)$.But even admitting a very large, and probably unrealistic,
value of $R^2_L(pp)$, say up to about $4fm^2$ the data would indicate a
presence of expansion even in collisions like p+Pb.Assuming a particular 
value of $R^2_L(pp)$ we can roughly estimate the expansion times as above.
For $R_L(pp)=1fm$ and $R_L(Pb+Pb) \approx 6fm$ we obtain $t_{expan} 
\approx 12.6fm/c$ for Pb-Pb interactions
 and with $R_L(S+Pb) \approx 4fm$ we have $t_{expan} \approx
8.4fm/c$ for S-Pb collisions.
\section{Comments and conclusions}
\label{comments}
Data on $R_L$ obtained in ion-ion collisions at the CERN-SPS [11,12,13]
give a valuable information on the space- time evolution of these
collisions.We have tried above to analyze these data by using a very
simple picture of ion- ion collision as a sum of nucleon- nucleon
interactions.The picture leads to a simple formula Eq.(10).We have used
this formula in two approaches.In the former we have introduced $R_L(pp)$
and the constant $c$ in Eq.(12) as free parameters.In this way data with
incident light ions can be interpreted as due to nuclear geometry, but
$R_L$ for Pb-Pb collisions is definitely higher than a geometrical model
can accomodate what gives evidence for the presence of collective 
expansion of matter in Pb-Pb collisions.The case of S-Pb and S-Au
collisions is difficult to interpret since there is an unexpectedly
large difference between the two data.It turns out also that an accurate
information on $R_L(pp)$ is of primary importance.

In the latter approach we have calculated the contribution of nuclear
geometry to $R_L(pp)$ in a more rigid way,leaving $R_L(pp)$ as the 
only unknown parameter.The resulting $R_L(AB)$
as given by geometry is almost independent
of $A,B$ and practically equal to $R_L(pp)$.The value of $R_L(pp)$ is thus
a key to interpreting data on $R_L$ for light- ion induced interactions.
Pb-Pb data are again much higher than geometry can explain.Our estimates
of the expansion time in Pb-Pb collision give $t_{expan} \approx 9fm/c$.

Study of the dependence of $R_L$ on A,B and on incident energy can, in our
opinion, reveal the changing dynamics of ion- ion collisions.This concerns
in particular presence of "thresholds" corresponding to changes of the
dynamical regime.

The absence of accurate information on $R_L(pp)$ and rather large
errors of  data make it difficult to see whether  data indicate
 presence of one or two "thresholds".The first one might be rather
low - somewhere
  between pp and S-C collisions,but this question depends crucially
on the value of $R_L(pp)$, and the second one between O-Au and Pb-Pb.
The existence of the latter seems very probable.  

\vspace*{2cm}     
{\bf Acknowledgements}The authors are indebted to C.Fabjan,D.Ferenc,
R.Lietava,G.Roche and B.Tom\'a\v{s}ik  for valuable discussions,
 to Yiota Foka for having organized a very stimulating HBT-Forum at
CERN in March 1996 and to Daniel Ferenc for calling our attention to
to an erroneous factor of two in some of equations.
 One of the authors (J.P.) is indebted to Guy Roche 
and Bernard Michel for hospitality at the Laboratoire de Physique
Corpusculaire,Universit\'e Blaise Pascal,Clermont-Ferrand.

\vspace*{1cm}
{\bf Figure Caption}
The dependence of $y={R_L}^2(A,B)$ as given in Refs.[11-14] on
$x=((A-1)^{1/3}+(B-1)^{1/3})^2$.Straight line fit corresponds to
$y=a+bx$ with $a=3fm^2$ and $b=0.13fm^2$.In making the fit we have excluded
data on $S+Pb$ and $Pb+Pb$ collisions.Note the $Pb+Pb$ point in the right
upper corner.For discussion of the second approach to data the reader
is invited to draw a line parallel with the x-axis and corresponding to
his preferred value of $R_L(pp)$.

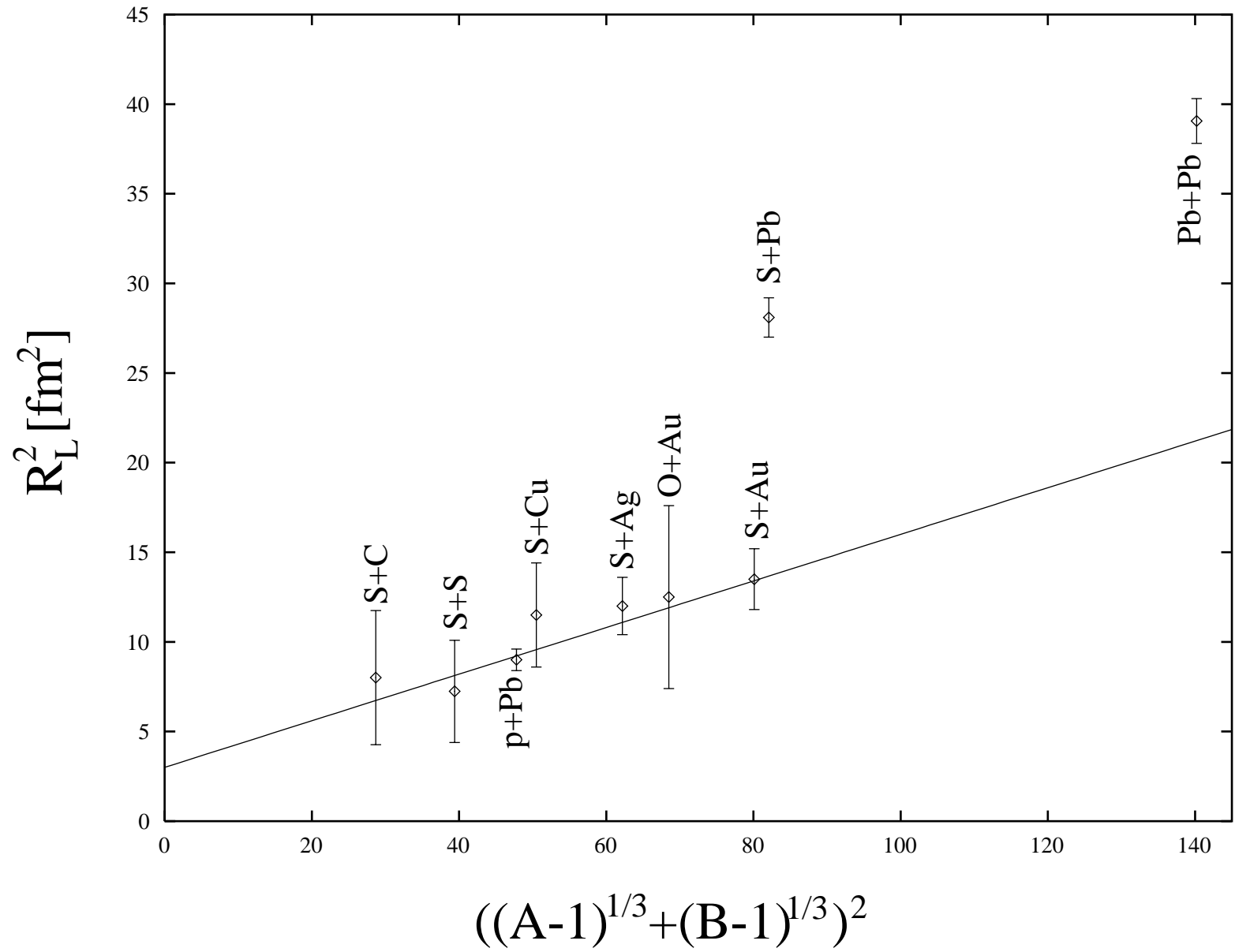
\end{document}